\documentclass[11pt,a4paper,english,superscriptaddress,aps,preprint,amsmath,amssymb]{revtex4}
\usepackage{graphicx}
\usepackage{graphics}
\usepackage{amsmath}
\usepackage{dcolumn}
\usepackage{amssymb}
\usepackage{bm}
\usepackage[latin1]{inputenc}

\begin{document}
\title{Gravitational Geometric Phase in the Presence of Torsion}
\author{Knut Bakke, Claudio Furtado and J. R. Nascimento}
\email{kbakke@fisica.ufpb.br,furtado@fisica.ufpb.br} 
\affiliation{Departamento de F\'{\i}sica, Universidade Federal da Para\'{\i}ba, Caixa Postal 5008, 58051-970, Jo\~ao Pessoa, Pb, Brazil}

\begin{abstract}
 We investigate the relativistic and non-relativistic  quantum dynamics of a neutral spin-$1/2$ particle submitted an external electromagnetic field in the presence of a cosmic dislocation.
We analyze the explicit contribution of the torsion in the geometric phase acquired in  the dynamic of  this neutral spinorial particle. We discuss the influence of the torsion in the relativistic geometric phase. Using the Foldy-Wouthuysen approximation, the non-relativistic quantum dynamics are studied and the influence of the torsion in the Aharonov-Casher and He-McKellar-Wilkens effects are discussed. 
\end{abstract}
\maketitle

\section{Introduction}
$ $ 

The study of gravity with torsion is a field of large interest to the physicists since a long time \cite{datta,hehl,5,1,rumpf}. A interesting review of the general relativity with spin and torsion has been done in  \cite{5}. The investigations  of fermions  interacting with  torsion also has been considered in the Ref.\cite{7}.  The mathematical apparatus for the Dirac equation coupled with torsion, in curved spacetime, is developed in \cite{1}. The formulation of the conservation laws
 in the spacetime with curvature and torsion was verified in \cite{8} and, the non-relativistic limit of the fermions coupled to torsion in the flat spacetime was studied in \cite{7}. Recently, some  papers has studied  the some possible 
physical effects of the torsion\cite{bagrov, hamon,doba,sing,shapiro}.   The physical effect of coupling of space-time torsion with fermios was theoretically  investigated, on the energy levels of atoms in a Hughes-Drever type experiments,  by L\"ammerzah\cite{3}.

Is well knowed that   a crystal with defects may in the continuum picture be described
as a space with curvature and torsion\cite{kroner,bilby,kleinert,anp:dia,6,katanaev92}. In this formulation we use the techniques of differential geometry to describe 
the strain and stress induced by the defect in an elastic medium. All this 
information is contained in the geometric quantities (metric, curvature tensor, 
etc.) that describe the elastic medium with defects. The boundary conditions 
imposed by the defect, in the elastic continuum, are accounted for by a non-Euclidean metric.  In the continuum limit, the solid can be viewed as a Riemann-Cartan manifold. In general, the defect corresponds to a singular curvature or 
torsion (or both) along the  defect line~\cite{katanaev92,furt}, where the curvature 
and the torsion of the manifold are associated to the topological defects, 
disclinations and dislocations, respectively. In this way, the quantum dynamics of electrons/holes in a crystal can be modelled by quantum dynamics of a particle in space with curvature and torsion.  
The influence of the torsion of the space in the dynamics of nonrelativistic particle has attracted attention of the physicists for a long time\cite{6,2,kaw,epl1,pla,furtpla1,furtepl1}.
Aurell\cite{2}  have studied the influence of torsion in the motion of conducting electrons in a quantum dot  and have discussed the physical implications of the torsion in classical as on non-relativistic quantum dynamics.

The study of the influence of external inertial and gravitational fields in the quantum dynamics of particles is a topic of great interest in the investigation of the structure of spacetime. In recent years, the concept on Berry's\cite{ber} phase was extended to consider the influence of  weak gravitational  and inertial fields \cite{papiprl,papinibook}. The  Relativistic generalization of geometric phase   was obtained  by Cai and Papini \cite{caimpl,caicqg} using the proper time method. Corichi and Pierri \cite{cori}  investigated the phase acquired by the quantum scalar particle when  transported along a closed path surrounding a rotating cosmic string. Mostafazadeh \cite{Mosta} employing the two-component formalism also investigated the appearance of a Berry's phase for a scalar particle in the background of a rotating cosmic string. Recently,  the Berry's quantum phase for a scalar particle and spinorial in the presence of a magnetic chiral cosmic string was investigated in Refs. \cite{furtassi,josevi}. Many authors have analyzed the Berry's phase in a series of  curved spaces \cite{feng,jil}. Recently,  the appearance of relativistic geometrical phase in the  dynamics of a neutral particles with permanents electric and magnetic dipoles  was investigated in a flat spacetime, without torsion by Anandan\cite{12,13,anandan}. In \cite{fur2} the noncommutativity version of geometric phase acquired by neutral particles was investigated. In the context of only the curved spacetime, as the dynamics as the appearance of relativistic phase of the same neutral particle was studied in \cite{bf1}. The geometric phase in flat space-time with torsion was studied in the reference \cite{shir} in the context of Einstein-Cartan model. The Berry phase of oscillating neutrinos was calculated  in Ref. \cite{capo} in the framework of theories of gravity with torsion. In this paper we are analyzed the influence of the torsion and the curvature of the spacetime in the  relativistic and nonrelativistic quantum phase acquired by wave function of a neutral  spin-$1/2$ particle with permanent electric and/or magnetic dipole moment in the presence of a electric and magnetic fields.  In the present contribution we also analyze the gravitational Aharonov-Casher\cite{17} and He-McKellar-Wilkens\cite{18,19} geometric phase in this dynamics.

The structure of the paper is given as: In section II, we give the general behaviour of the neutral particles with permanent dipoles in the curved spacetime background with torsion. In section III, we present the geometrical aspects of the topological defect. In section IV, we work the Dirac equation and study the relativistic phase shift acquired by the wave function of the neutral particles. In section V, we investigate the quantum dynamics of this particle in the non-relativistic limit. In section VI, we study the interference effects caused by the curvature and torsion. At the end, in section VII, we present the conclusions.

\section{ Spin-$1/2$ Neutral  Particles in a Curved Space-time with Torsion}
$ $ 
In this section we consider the Dirac equation for a neutral spin-half particle with nonzero magnetic and electric dipoles.
The relativistic quantum dynamics of a single neutral spin-half particle with nonzero magnetic and electric dipoles that we shall consider in this section is when the particle is moving in a external electromagnetic field in the curved space-time with torsion. (where we used $\hbar=c=1$). The equation which describes this is   
\begin{eqnarray}
i\gamma^{\mu}\,\nabla_{\mu}\psi+\frac{\mu}{2}\Sigma^{\mu\nu}\,F_{\mu\nu}\psi-i\frac{d}{2}\,\Sigma^{\mu\nu}\,\gamma^{5}\,F_{\mu\nu}-m\psi=0,
\label{1.2}
\end{eqnarray}
where $\mu$ is the magnetic dipole moment, $d$ is the electric dipole moment, the Greek index $(\mu,\nu)$ indicates the spacetime indices. The tensor $F_{\mu\nu}$ is  
\begin{eqnarray}
F_{\mu\nu}&=&\left\{\vec{E},\vec{B}\right\},\,\,\,\,\,F_{\mu\nu}=-F_{\nu\mu} \nonumber\\
F_{0\alpha}&=&E_{\alpha};\,\,\,\,\,F_{\alpha\beta}=-\epsilon_{\alpha\beta\gamma}\,B^{\gamma}.
\label{1.8}
\end{eqnarray}
with $(\alpha,\beta,\gamma)$ being the spatial indices of the spacetime. In curved spacetime we need to define a local frame where we can study the electromagnetic effects in the presence of a gravitational field. We define our local frame through the tetrads $e^{a}_{\,\,\,\mu}\left(x\right)$ which are the components of the non-coordinate basis $\hat{\Theta}^{a}=e^{a}_{\,\,\,\mu}\,dx^{\mu}$ \cite{naka} and its inverse $dx^{\mu}=e^{\mu}_{\,\,\,a}\,\hat{\Theta}^{a}$. The covariant derivative in the curved spacetime with torsion is given by
\begin{eqnarray}
\nabla_{\mu}&=&\partial_{\mu}+\Gamma_{\mu}+K_{\mu}\nonumber\\
&=&\partial_{\mu}-\frac{i}{4}\,\omega_{\mu ab}\Sigma^{ab}+\frac{i}{4}K_{\mu ab}\Sigma^{ab},
\label{1.3}
\end{eqnarray}
where $\Sigma^{ab}=\frac{i}{2}\left[\gamma^{a},\gamma^{b}\right]$ and the indices $(a,b,c=0,1,2,3)$ indicate the local reference frame. The $\gamma^{a}$ matrices are defined in the local reference frame and are identical to the Dirac matrices in the flat spacetime, \textit{i.e.},
\begin{eqnarray}
\gamma^{0}=\hat{\beta}=\left(
\begin{array}{cc}
1 & 0 \\
0 & -1 \\
\end{array}\right),\,\,\,\,\,\,
\gamma^{i}=\hat{\beta}\,\alpha^{i}=\left(
\begin{array}{cc}
 0 & \sigma^{i} \\
-\sigma^{i} & 0 \\
\end{array}\right),
\label{1.10}
\end{eqnarray}
with $\sigma^{i}$ being the Pauli matrices whose satisfy the relation $\left(\sigma^{i}\,\sigma^{j}+\sigma^{j}\,\sigma^{i}\right)=-2\,\eta^{ij}$, where $\eta^{ab}=diag(- + + +)$ is the Minkowsky tensor and the index $(i,j,k=1,2,3)$ are the spacial index of the local reference frame. The $\gamma^{5}$ matrix is defined as
\begin{eqnarray}
\gamma^{5}=\frac{-i}{24}\,\epsilon_{\mu\nu\alpha\beta}\,\gamma^{\mu}\,\gamma^{\nu}\,\gamma^{\alpha}\,\gamma^{\beta}=i\,\gamma^{0}\,\gamma^{1}\,\gamma^{2}\,\gamma^{3}=-\gamma_{5}=\left(
\begin{array}{cc}
0 & 1 \\
1 & 0 \\
\end{array}\right),
\label{1.11}
\end{eqnarray}
and, for convenience, we define the spin vector $\vec{\Sigma}$ as
\begin{eqnarray}
\vec{\Sigma}=\left(
\begin{array}{cc}
\vec{\sigma} & 0 \\
0 & \vec{\sigma} \\
\end{array}\right).
\label{1.12}
\end{eqnarray}

Back to the expression to the covariant derivative (\ref{1.3}), we have a one-form connection $\omega^{a}_{\,\,\,b}=\omega_{\mu\,\,\,\,b}^{\,\,\,a}\,dx^{\mu}$ and $K_{\mu ab}$ which is related with the contortion tensor by \cite{7}
\begin{eqnarray}
K_{\mu ab}= K_{\beta\nu\mu}\left(e^{\nu}_{\,\,\,a}\,e^{\beta}_{\,\,\,b}-e^{\nu}_{\,\,\,b}\,e^{\beta}_{\,\,\,a}\right),
\label{1.4}
\end{eqnarray}
and the contortion tensor is related with the torsion tensor as
\begin{eqnarray}
K^{\beta}_{\,\,\,\nu\mu}=\frac{1}{2}\left(T^{\beta}_{\,\,\,\nu\mu}-T_{\nu\,\,\,\,\mu}^{\,\,\,\beta}-T^{\,\,\,\beta}_{\mu\,\,\,\,\nu}\right).
\label{1.5}
\end{eqnarray}
We are using the definitions given in \cite{7} where the torsion tensor is antisymmetric in the last two indices while the contortion tensor is antisymmetric in the first two indices. It is usually convenient to write the torsion tensor into three irreducible components \cite{8,7}:
\begin{eqnarray}
T_{\mu}=T^{\beta}_{\,\,\,\mu\beta};\,\,\,\,\,\,S^{\alpha}=\epsilon^{\alpha\beta\nu\mu}\,T_{\beta\nu\mu};
\label{1.6}
\end{eqnarray}
and in the tensor $q_{\beta\nu\mu}$ which satisfy the conditions $q^{\beta}_{\,\,\mu\beta}=0$ and $\epsilon^{\alpha\beta\nu\mu}\,q_{\beta\nu\mu}=0$. So, the torsion tensor becomes
\begin{eqnarray}
T_{\beta\nu\mu}=\frac{1}{3}\left(T_{\nu}\,g_{\beta\mu}-T_{\mu}\,g_{\beta\nu}\right)-\frac{1}{6}\,\epsilon_{\beta\nu\mu\gamma}\,S^{\gamma}+q_{\beta\nu\mu}.
\label{1.7}
\end{eqnarray}

So, the expression (\ref{1.2}) for the Dirac equation becomes
\begin{eqnarray}
i\gamma^{a}\,e^{\mu}_{\,\,\,a}\,\nabla_{\mu}\psi+\frac{\mu}{2}\,F_{\mu\nu}\,e^{\mu}_{\,\,\,a}\,e^{\nu}_{\,\,\,b}\,\Sigma^{ab}\psi-i\frac{d}{2}\,\Sigma^{ab}\,e^{\mu}_{\,\,\,a}\,e^{\nu}_{\,\,\,b}\,\gamma^{5}\,F_{\mu\nu}-m\psi=0,
\label{1.9}
\end{eqnarray} 
With this definitions in ours hands, we are able to build the local reference frames for observers that are located in some place of the spacetime. We shall see this in the next section where we choose a general relativity background.
%%%%%%%%%%%%%%%%%%%%%%%%%%%%%%%%%%%%%%%%%%%%%%%%%%%%%%%%%%%%%%%%%%%%%%%%%%%%%%%%%%%%%%%%%%%%%%%%%%%%%%%%%%%%%%%%%%%%%%%%%%%%%
\section{The cosmic Dislocation}
%%%%%%%%%%%%%%%%%%%%%%%%%%%%%%%%%%%%%%%%%%%%%%%%%%%%%%%%%%%%%%%%%%%%%%%%%%%%%%%%%%%%%%%%%%%%%%%%%%%%%%%%%%%%%%%%%%%%%%%%%%%%%%%%%
$ $ 

In this section we develop the structure of the curved spacetime that we work out along this paper. We choose a topological defect called Cosmic dislocation\cite{letel,gal} which its line element is given by
\begin{eqnarray}
ds^{2}=-dt^{2}+d\rho^{2}+\eta^{2}\rho^{2}d\varphi^{2}+\left(dz-\chi\,d\varphi\right)^{2},
\label{2.1}
\end{eqnarray}
where $\eta=1-4G\nu$ is the angular deficit defined at the range $0<\eta<1$, with $\nu$ being the linear mass density. The parameter $\chi$ is related to the torsion of the defect. It is related with the Burges vector if we consider the crystallography language. The azimuthal angle varies in the interval: $0\leq\varphi<2\pi$. The deficit angle can assume only values in which $\eta<1$ (unlike of this, in \cite{katanaev92,furt}, it can assume values greater than 1, which correspond to an anti-conical space-time with negative curvature). This geometry possess a conical singularity represented by the following curvature tensor
\begin{eqnarray}
\label{curv}\label{curva}
R_{\rho,\varphi}^{\rho,\varphi}=\frac{1-\eta}{4\eta}\delta_{2}(\vec{r}),
\end{eqnarray}
where $\delta_{2}(\vec{r})$ is the two-dimensional delta function. This behavior of the curvature tensor is denominated conical singularity~\cite{staro}. The conical singularity gives rise to the curvature concentrated on the cosmic string axis, in all  other places the curvature is null.

It is convenient to construct a local reference frame where we can define the spinors in curved spacetime. We can build the local reference frame through a non-coordinate basis $\hat{\theta}^{a}=e^{a}_{\,\,\,\mu}\,dx^{\mu}$, which its components $e^{a}_{\,\,\,\mu}\left(x\right)$ and satisfy the following relation \cite{naka}
\begin{eqnarray}
g_{\mu\nu}\left(x\right)=e^{a}_{\,\,\,\mu}\left(x\right)\,e^{b}_{\,\,\,\nu}\left(x\right)\,\eta_{ab}.
\label{2.2}
\end{eqnarray}
the components of the non-coordinate basis $e^{a}_{\,\,\,\mu}\left(x\right)$ are called \textit{tetrads or Vierbein} and them form our local reference frame. The tetrads has a inverse define as $dx^{\mu}=e^{\mu}_{\,\,\,a}\,\hat{\theta}^{a}$, where 
\begin{eqnarray}
e^{a}_{\,\,\,\mu}\,e^{\mu}_{\,\,\,b}=\delta^{a}_{\,\,\,b}\,\,\,\,\,\,\,e^{\mu}_{\,\,\,a}\,e^{a}_{\,\,\,\nu}=\delta^{\mu}_{\,\,\,\nu}.
\label{2.3}
\end{eqnarray}

We have many forms to construct the tetrads, so, we choose in this system that our tetrads are given by
\begin{eqnarray}
\hat{\theta}^{0}&=&dt\nonumber\\
\hat{\theta}^{1}&=&\cos\varphi\,d\rho-\eta\rho\sin\varphi\,d\varphi\nonumber\\
\hat{\theta}^{2}&=&\sin\varphi\,d\rho+\eta\rho\cos\varphi\,d\varphi\nonumber\\
\hat{\theta}^{3}&=&dz-\chi\,d\varphi.
\label{2.4}
\end{eqnarray}
The matricial form of the tetrads and its inverse is
\begin{eqnarray}
e^{a}_{\,\,\,\mu}\left(x\right)=\left(
\begin{array}{cccc}
1 & 0 & 0 & 0 \\
0 & \cos\varphi & -\eta\rho\sin\varphi & 0 \\
0 & \sin\varphi & \eta\rho\cos\varphi & 0 \\
0 & 0 & \chi & 1 \\
\end{array}\right),\,\,\,
e^{\mu}_{\,\,\,a}=\left(
\begin{array}{cccc}
1 & 0 & 0 & 0 \\
0 & \cos\varphi & \sin\varphi & 0 \\
0 & -\frac{\sin\varphi}{\eta\rho} & \frac{\cos\varphi}{\eta\rho} & 0 \\
0 & \frac{\chi}{\eta\rho}\sin\varphi & -\frac{\chi}{\eta\rho}\cos\varphi & 1 \\
\end{array}\right).
\end{eqnarray}

With the information about the choice of the local reference frame, we can obtain the one-form connection $\omega^{a}_{\,\,\,b}=\omega_{\mu\,\,\,\,b}^{\,\,\,a}\,dx^{\mu}$ and the components of the torsion tensor through the Cartan's structure equation \cite{naka}
\begin{eqnarray}
T^{a}=d\hat{\theta}^{a}+\omega^{a}_{\,\,\,b}\wedge\hat{\theta}^{b},
\label{2.5}
\end{eqnarray}
where the operator $d$ is the exterior derivative and the symbol $\wedge$ means the exterior product. Observing the symmetry of the defect, we obtain
\begin{eqnarray}
\omega_{\varphi\,\,\,2}^{\,\,\,1}=-\omega_{\varphi\,\,\,1}^{\,\,\,2}=1-\eta.
\label{2.6}
\end{eqnarray}
and
\begin{eqnarray}
T^{3}&=&\chi\,\delta\left(\rho\right)\,d\varphi\wedge d\rho,
\label{2.7}
\end{eqnarray}
where $\delta\left(\rho\right)$ is a delta function in spacetime indices. With this, we have one contribution of the torsion given in the singularity of the defect. 

%%%%%%%%%%%%%%%%%%%%%%%%%%%%%%%%%%%%%%%%%%%%%%%%%%%%%%%%%%%%%%%%%%%%%%%%%%%%%%%%%%%%%%%%%%%%%%%%%%%%%%%%%%%%%%%%%%%%%%%%%%%%%%%%%%
\section{Dirac equation in cosmic dislocation background }
%%%%%%%%%%%%%%%%%%%%%%%%%%%%%%%%%%%%%%%%%%%%%%%%%%%%%%%%%%%%%%%%%%%%%%%%%%%%%%%%%%%%%%%%%%%%%%%%%%%%%%%%%%%%%%%%%%%%%%%%%%%%%%%%%%%
In \cite{7}, the non-relativistic limit of Dirac equation for spin half particle in the presence of external electromagnetic and torsion field was considered. Thus, the Pauli equations was obtained in the frame work of the Foldy-Wouthuysen transformation.The Dirac equation that describes a spin-$1/2$  neutral particle with non-zero magnetic and electric dipole moments moving in an external electromagnetic field is given by the expressions (\ref{1.2}) or (\ref{1.9}).
Using the expressions (\ref{1.3}), (\ref{2.6}) and (\ref{2.7}) the first part of the spinorial connection, $\Gamma_{\mu}$, has only one non-zero component
\begin{eqnarray}
\Gamma_{\varphi}&=&\frac{1}{4}\left(1-\eta\right)\left[\gamma^{1},\gamma^{2}\right]=
-\frac{i}{2}\left(1-\eta\right)\,\Sigma^{3}.
\label{3.1}
\end{eqnarray}

Thus, using the expressions (\ref{1.4}), (\ref{1.5}) and (\ref{1.7}) in the second part of the spinorial connection $K_{\mu}$, the Dirac equation in curved spacetime with torsion (\ref{1.2}) becomes 
\begin{eqnarray}
i\,\gamma^{t}\,\frac{\partial\psi}{\partial t}&+&i\,\gamma^{\rho}\left(\partial_{\rho}-\frac{1}{2}\frac{\left(1-\eta\right)}{\eta\rho}-\mu\,\hat{\beta}\,E_{\rho}+d\,\hat{\beta}\,B_{\rho}\right)\psi+i\frac{\gamma^{\varphi}}{\eta\rho}\,\frac{\partial\psi}{\partial\varphi}+i\,\gamma^{z}\,\frac{\partial\psi}{\partial z}\nonumber\\
&-&\frac{1}{8}\vec{\Sigma}\cdot\vec{S}\,\psi-\mu\,\vec{\Sigma}\cdot\vec{B}\,\psi-d\,\vec{\Sigma}\cdot\vec{E}\,\psi+\frac{1}{8}\,\hat{\beta}\,S^{0}\,\gamma^{5}\,\psi-m\psi=0.
\label{3.2}
\end{eqnarray}
In this background the matrices $\gamma^{\mu}=e^{\mu}_{\,\,\,a}\,\gamma^{a}$ are given by
\begin{eqnarray}
\gamma^{t}&=&e^{t}_{\,\,\,a}\,\gamma^{a}=\gamma^{0};\\
\gamma^{\rho}&=&e^{\rho}_{\,\,\,a}\,\gamma^{a}=\cos\varphi\,\gamma^{1}+\sin\varphi\,\gamma^{2};\\
\gamma^{\varphi}&=&e^{\varphi}_{\,\,\,a}\,\gamma^{a}=-\sin\varphi\,\gamma^{1}+\cos\varphi\,\gamma^{2};\\
\gamma^{z}&=&e^{z}_{\,\,\,a}\,\gamma^{a}=\gamma^{3}.
\label{3.3}
\end{eqnarray}
Note that  Eq.(\ref{3.2})  terms that depend explicitly of the torsion of the space-time. Notice that, in the term $\frac{1}{8}\vec{\Sigma}\cdot\vec{S}$, the torsion play the role   similar to  the coupling of magnetic field with the spin  in the flat case.
Now, let us discuss the relativistic geometric phase in this dynamics. We consider that the spinor $\psi$ can be written in the following form
\begin{equation}\label{phased}
\psi=e^{i\phi}\psi_{0}
\end{equation}
where $\psi_{0}$ is phase and is the solution of Dirac equation in the absence of fields and $\phi$ is a phase. Substituting  Eq.(\ref{phased}) in  Eq. (\ref{3.2}), we obtain a relativistic phase given by the expression 
\begin{eqnarray}\label{relatpha}
\phi_{R} =\oint \left(\frac{1}{2}\left(1-\eta\right)\,\Sigma^{3}-\mu\,\hat{\beta} \,\left(\vec{\Sigma}\times\vec{E}\right)_{\varphi}+ d\,\hat{\beta}\,\left(\vec{\Sigma}\times\vec{B}\right)_{\varphi}+\frac{1}{8}\,e^{i}_{\,\,\,\varphi}\,S^{0}\,\Sigma_{i}\right)\,d\varphi,
\end{eqnarray}
Note that this relativistic quantum phase has four contributions, where the first three contributions was discussed in \cite{bf1}. The last term of (\ref{relatpha}) comes from the torsion field and show us how the torsion field are present in the relativistic quantum phase. Using the expression (\ref{1.6}), we have that this contribution indicates a flux of the torsion field as indicated in \cite{anandan}
\begin{eqnarray}
\phi_{T}&=&\frac{1}{8}\oint\,e^{i}_{\,\,\,\varphi}\,S^{0}\,\Sigma_{i}\,d\varphi=-\frac{\pi}{2}\,\chi\,\Sigma^{3}.
\end{eqnarray}
This term represent the influence of the torsion in the relativistic geometric phase. This effect can be investigated in the interferometric experiment , where the defect is circulated by the interferometric path. This torsion contribution to the geometric phase is topological in the sense that this not depend of the path of the particle encircle the defect and depend the winding number.
Again, we must observe that the first term in equation (\ref{relatpha}) is the relativistic Berry's geometric phase which was proposed in \cite{caimpl,caicqg} for spin $1/2$ particle using the weak field approximation to the curved spacetime. As discussed in \cite{bf1}, we can write this term in the following form
\begin{equation}\label{pap}
 \phi_{p}= \frac{1}{4}\oint R_{\mu \nu \delta \lambda} J^{ \delta \lambda}d\tau^{\mu \nu},
\end{equation}
where $J^{ \delta \lambda}= L^{ \delta \lambda} + \Sigma^{ \delta \lambda}$ is the total angular momentum of the particle. Taking the curvature tensor in the expression (\ref{curva}), we find that (\ref{pap}) is
\begin{equation}
\phi_{p}=\oint \frac{1}{2}\left(1-\eta\right)\,\Sigma^{3}  d\varphi.
\end{equation}

We emphasize that in our study, we do not use this approximation, and obtain that the phase found in \cite{caimpl,caicqg} is generic. 

To end this section, we argue that the two others contributions in (\ref{relatpha}) are due to the magnetic  and electric dipole moments to relativistic Anandan geometric phase \cite{12,13} in the presence of cosmic string, but now with the presence of torsion.  they are given by
\begin{equation}\label{relatpha1}
\phi =\oint \left(- \mu \,\left(\vec{\Sigma}\times\vec{E}\right)_{\varphi}+d \,\left(\vec{\Sigma}\times\vec{B}\right)_{\varphi}\right)\,d\varphi,
\end{equation}
If we take the limit $\eta\longrightarrow 1$ and $\chi=0$, we obtain flat spacetime results for the Anandan geometric phase.

%%%%%%%%%%%%%%%%%%%%%%%%%%%%%%%%%%%%%%%%%%%%%%%%%%%%%%%%%%%%%%%%%%%%%%%%%%%%%%%%%%%%%%%%%%%%%%%%%%%%%%%%%%%%%%%%%%%%%%%%%%%%
\section{the Foldy-Wouthuysen approximation}
%%%%%%%%%%%%%%%%%%%%%%%%%%%%%%%%%%%%%%%%%%%%%%%%%%%%%%%%%%%%%%%%%%%%%%%%%%%%%%%%%%%%%%%%%%%%%%%%%%%%%%%%%%%%%%%%%%%%%%%%%%%

In this section we wish to write the Dirac equation for the neutral particle with permanent electric and magnetic moment of dipoles in the topological defect background when there are the influence of an external electric and magnetic fields. In this way, we study the non-relativistic limit of the Dirac equation using the Foldy-Worthuysen approximation. We shall start writing the Dirac equation in the form 
\begin{eqnarray}
i\frac{\partial\psi}{\partial t}=H\psi.
\label{4.1}
\end{eqnarray}
So, taking the equation (\ref{1.9}), we can rewrite as 
\begin{eqnarray}
i\frac{\partial\psi}{\partial t}&=&m\hat{\beta}\psi+\vec{\alpha}\cdot\vec{p}\psi-i\vec{\alpha}\cdot\vec{\xi}+\frac{1}{8}\vec{\Sigma}\cdot\vec{S}\,\psi+\frac{1}{8}S_{0}\,\gamma^{5}\,\psi\nonumber\\
&+&i\mu\hat{\beta}\left(-\vec{\alpha}\cdot\vec{E}+\hat{\beta}\,\vec{\Sigma}\cdot\vec{B}\right)\psi+id\hat{\beta}\left(-i\vec{\Sigma}\cdot\vec{E}+\vec{\alpha}\cdot\vec{B}\right)\psi, 
\label{4.2}
\end{eqnarray}
where we call, in the local reference frame, $p_{i}=-i\,e^{\alpha}_{\,\,\,i}\,\partial_{\alpha}$; $E_{i}=e^{\alpha}_{\,\,\,i}\,E_{\alpha}$; $B_{i}=e^{\alpha}_{\,\,\,i}\,B_{\alpha}$ and where we define in \cite{bf1} the quantity
\begin{eqnarray}
\xi_{i}=\frac{1}{4}e^{\alpha}_{\,\,\,i}\,\omega_{\alpha jk}\,\Sigma^{jk}=-\frac{i}{2}\left(1-\alpha\right)\Sigma^{3}\,e^{\varphi}_{\,\,\,i},
\label{4.3}
\end{eqnarray}
and  
\begin{eqnarray}
\vec{\pi}=\vec{p}+i\mu\hat{\beta}\,\vec{E}-id\hat{\beta}\,\vec{B}-i\vec{\xi},
\label{4.4}
\end{eqnarray}
where it is defined in the local reference frame and it is identical to that done in \cite{fur2} in the flat spacetime and \cite{bf1} in the presence of the topological  defect. The Dirac equation (\ref{4.2}) becomes
\begin{eqnarray}
i\frac{\partial\psi}{\partial t}=m\hat{\beta}\psi+\vec{\alpha}\cdot\vec{\pi}\,\psi+\frac{1}{8}\vec{\Sigma}\cdot\vec{S}\,\psi+\frac{1}{8}S_{0}\,\gamma^{5}\,\psi+\mu\,\hat{\beta}\,\vec{\Sigma}\cdot\vec{B}\psi+d\,\hat{\beta}\,\vec{\Sigma}\cdot\vec{E}\,\psi.
\label{4.5}
\end{eqnarray}
Now, we apply the Foldy-Wouthuysen approximation \cite{fw} in the Dirac equation. This approach permit us to investigate the non-relativistic limit of the Dirac equation in the curved spacetime with torsion.  The Hamiltonian of the system must be write as a linear combination of even terms $\hat{\epsilon}$ and odd terms $\hat{O}$ as following
\begin{eqnarray}
H=\hat{\beta}\,m+\hat{O}+\hat{\epsilon},
\label{4.6}
\end{eqnarray}
where the even and odd operators need to be hermitian operators and must satisfy the relations
\begin{eqnarray}
\begin{array}{c}
\hat{O}\,\hat{\beta}+\hat{\beta}\,\hat{O}=0,\\
\hat{\epsilon}\,\hat{\beta}-\hat{\beta}\,\hat{\epsilon}=0.\\
\end{array}
\label{4.7}
\end{eqnarray}

In the final result of this approximation we consider just the terms up to order of $m^{-1}$. So, the Hamiltonian of the system in the low-energy limit becomes
\begin{eqnarray}
H'=\hat{\beta}\,m+\frac{\hat{\beta}}{2m}\hat{O}^{2}+\hat{\epsilon}.
\label{4.8}
\end{eqnarray}  
Considering the expression (\ref{4.5}) we have that 
\begin{eqnarray}
\hat{O}&=&\vec{\alpha}\cdot\vec{\pi}+\frac{1}{8}S_{0}\,\gamma^{5}\\
\hat{\epsilon}&=&\mu\,\hat{\beta}\,\vec{B}\cdot\vec{\Sigma}+d\,\hat{\beta}\,\vec{E}\cdot\vec{\Sigma}+\frac{1}{8}\vec{\Sigma}\cdot\vec{S},
\label{4.9}
\end{eqnarray}
and the expression for the Hamiltonian (\ref{4.8}) becomes
\begin{eqnarray}
H'&=&\hat{\beta}\,m+\frac{\hat{\beta}}{2m}\left(\vec{p}+\vec{\Xi}\right)^{2}-\frac{\mu^{2}\,E^{2}}{2m}-\frac{d^{2}\,B^{2}}{2m}+\frac{1}{8}\vec{\Sigma}\cdot\vec{S}\nonumber\\
&+&\frac{\mu}{2m}\vec{\nabla}\cdot\vec{E}+\frac{d}{2m}\vec{\nabla}\cdot\vec{B}+d\,\hat{\beta}\,\vec{\Sigma}\cdot\vec{E}+\mu\,\hat{\beta}\,\vec{\Sigma}\cdot\vec{B}
\label{4.10}
\end{eqnarray}
where we have introduced  the vector $\vec{\Xi}$ as  in \cite{bf1}, but with a new contribution due the torsion field
\begin{eqnarray}
\vec{\Xi}=\mu\,\hat{\beta}\,\vec{\Sigma}\times\vec{E}-d\,\hat{\beta}\,\vec{\Sigma}\times\vec{B}+i\vec{\xi}+\frac{1}{8}\vec{\Sigma}\,S^{0}.
\label{4.11}
\end{eqnarray}

The behaviour of the electric and magnetic dipoles in the presence of an angular deficit and torsion where there exists the action of external electric and magnetic fields is described by the Hamiltonian (\ref{4.10}). The topology of the defect (\ref{2.1}) is evident due the last two terms in the expression (\ref{4.11}). We can see that if we consider that $\chi=0$, \textit{i.e.}, in the absence of torsion, we recuperate the results obtained by the authors in \cite{bf1} in the cosmic string background without torsion. However, if we also consider $\eta\rightarrow 1$, we have the same configuration obtained by Passos, \textit{at all} in the flat spacetime \cite{fur2}. The effects due the configuration of the dipoles inside a topological defect with the influence of the external fields will be discussed in the next section. Notice that the term $\frac{1}{8}\vec{\Sigma}\cdot\vec{S}$ in Eq. (\ref{4.10}) give the coupling of the torsion with the spin of the particle in the non-relativistic limit. The  physical effects of this contribution can be investigated in condensed matter sytems where the torsion play a important role.

%%%%%%%%%%%%%%%%%%%%%%%%%%%%%%%%%%%%%%%%%%%%%%%%%%%%%%%%%%%%%%%%%%%%%%%%%%%%%%%%%%%%%%%%%%%%%%%%%%%%%%%%%%%%%%%%%%%%%%%%%%%%%%%%%%
\section{Gravitational Nonrelativistic Geometric Phase and Torsion}
%%%%%%%%%%%%%%%%%%%%%%%%%%%%%%%%%%%%%%%%%%%%%%%%%%%%%%%%%%%%%%%%%%%%%%%%%%%%%%%%%%%%%%%%%%%%%%%%%%%%%%%%%%%%%%%%%%%%%%%%%%%%%%%%%

We start this section writing the Schr\"ondiger equation considering  the terms that contributes to the appearance of the geometric quantum phases in the wave function. This becomes clear when we consider that the charge densities  are concentrated on the symmetry axis of a cosmic dislocation, with the dipoles oriented along the $z$-axis. In that way, the fields are cylindrically radial and are given by
\begin{eqnarray}
\vec{E}=\frac{\lambda_{e}}{\rho}\hat{\rho};\,\,\,\,\vec{B}=\frac{\lambda_{m}}{\rho}\hat{\rho},
\label{5.4}
\end{eqnarray}
where $\lambda_{e}$ is and electric charge density that creates the electric field and $\lambda_{m}$ is a constant that depends on the arrays of magnetic dipoles that generate a magnetic field. Note that the index of the fields indicates the general coordinate system, thus, we need to transform this configuration of fields to the configuration of fields in the local reference frame.

Hence, with this configuration of the external fields, the last four terms of the Hamiltionian (\ref{4.10}) give no contribution to the geometric phase acquired by the wave function. The terms proportional to $E^{2}$ and $B^{2}$ also do not contribute to the geometric phase because they are local terms \cite{12,13,fur2}. Thus, taking the remaining terms of the Hamiltonian (\ref{4.10}), the Schr\"odinger equation becomes
\begin{eqnarray}
\frac{1}{2m}\left(\vec{p}+\mu\,\hat{\beta}\,\vec{\Sigma}\times\vec{E}-d\,\hat{\beta}\,\vec{\Sigma}\times\vec{B}+i\vec{\xi}+\frac{1}{8}S_{0}\vec{\Sigma}\right)^{2}\,\Psi-\frac{\mu^{2}\,E^{2}}{8m}\,\Psi-\frac{d^{2}\,B^{2}}{8m}\,\Psi=E\,\Psi.
\label{5.1}
\end{eqnarray}   
The quantum phase can be obtained if we consider the ansatz:
\begin{eqnarray}
\Psi=e^{i\Phi}\,\psi,
\label{5.2}
\end{eqnarray}
where $\psi$ is the solution of the equation
\begin{eqnarray}
\frac{1}{2m}\,p^{2}\psi-\frac{\mu^{2}\,E^{2}}{8m}\,\psi-\frac{d^{2}\,B^{2}}{8m}\,\psi=E\,\psi.
\label{5.3}
\end{eqnarray}

If we consider that no charges are concentrated in the axis of symmetry of the defect, we have that a quantum phase given by the last two terms of the expression (\ref{4.11}), i.e. using (\ref{1.6}) in the contribution given by $S_{0}$ we have,
\begin{eqnarray}
\Phi&=&\oint\,e^{i}_{\,\,\,\mu}\left(i\xi_{i}\right)dx^{\mu}-\frac{1}{8}\sigma^{3}\oint\,T^{3}_{\,\,\,\mu\nu}dx^{\mu}\wedge dx^{\nu}\nonumber\\
&=&\left(1-\eta\right)\,\pi\,\sigma^{3}-\frac{\pi}{2}\,\chi\,\sigma^{3}.
\label{5.5}
\end{eqnarray} 
The first term give us the contribution of the angular deficit of the defect in the quantum phase. This term is identically to obtained by Furtado, \textit{at all} \cite{fur3} through the application of the parallel transport of the vector in a closed path in the graphene cones. The second term above explicit us the contribution of the torsion of the defect. 

Now, if we take into account the local reference frame and that the charges are concentrated on the symmetry axis of the topological defect, the quantum phase of this system is given by
\begin{eqnarray}
\Phi&=&\oint\,\Xi_{\mu}\,dx^{\mu}=\oint\,\Xi_{i}\,e^{i}_{\,\,\,\mu}\,dx^{\mu}=\int^{2\pi}_{0}\,\Xi_{i}\,e^{i}_{\,\,\,\varphi}\,d\varphi\nonumber\\
&=&\left(1-\eta\right)\,\pi\,\sigma^{3}-\frac{\pi}{2}\,\chi\,\sigma^{3}+\left(\mu\,\lambda_{e}-d\,\lambda_{m}\right)\,2\pi\,\sigma^{3},
\label{5.55}
\end{eqnarray}
where we considered the two-component spinor field. The contribution of the defect can be view clearly through the angular deficit parameter $\eta$ and the torsion parameter $\chi$ that appear in the first two terms of the expression (\ref{5.55}) and coupled with the dipoles moments. If we take $\eta\rightarrow 1$ and $\chi=0$, we recuperate the results obtained by Furtado and Lima Ribeiro \cite{fur1} in the absence of the topological defect. 

If we take $d=0$, we obtain the analogue effect of the Aharonov-Casher effect in the presence of topological defect. The quantum phase in this case becomes  
\begin{eqnarray}
\Phi_{AC}=\left(1-\eta\right)\,\pi\,\sigma^{3}-\frac{\pi}{2}\,\chi\,\sigma^{3}+2\pi\,\mu\,\lambda_{e}\,\sigma^{3}.
\label{5.6}
\end{eqnarray}

Our next step is take $\mu=0$ and obtain the analogue effect of the He-McKellar-Wilkens effect in the presence of the topological defect. The quantum phase becomes 
\begin{eqnarray}
\Phi_{HMW}=\left(1-\eta\right)\,\pi\,\sigma^{3}-\frac{\pi}{2}\,\chi\,\sigma^{3}-2\pi\,d\,\lambda_{m}\,\sigma^{3}.
\label{5.7}
\end{eqnarray}

Observe that if we take $\chi=0$ in the quantum phases (\ref{5.6}) and (\ref{5.7}) we recuperate the results given in \cite{bf1}. All the results obtained above (\ref{5.5}), (\ref{5.55}), (\ref{5.6}) and (\ref{5.7}) demonstrate the influence of the topological defect in the geometric quantum phase acquired by the wave function in the dynamics of the neutral particle. 

In that way, the general result given in the expression (\ref{5.5}) show us the geometric phase absorbs the topology of the defect in the dynamics of the neutral particle with a permanent electric and magnetic moment of dipole.

\section{Conclusions}
$ $ 
We investigated the influence of the torsion and the deficit of angle produced by a topological defect on the geometric phases acquired by the wave functions of relativistic neutral particles. The contribution of the torsion given here is equivalently to the flux of torsion pointed out by Anandan in \cite{anandan} and has its origin due to a disclination and dislocation if we treat in the context of the Volterra process in the theory of the elasticity. 

We investigated the non-relativistic quantum phase using the Foldy-Wouthuysen approximation. Again, the new contribution to the geometric phases was given by deficit of angle and torsion without assuming the adiabatic approximation or the weak field approximation. This geometric phase is identical to Anandan's geometric phase.

In both quantum phases, relativistic and non-relativistic, we can see that if we take the limit $\chi\rightarrow 1$ we are going to the flat spacetime, but the torsion continues with its influence on the quantum phase. If we take $\chi=0$ and $\eta\neq 1$ we have the result obtained by the authors in \cite{bf1}.

Finally, we investigated the effects analogue to the Aharonov-Casher and He-McKellar-Wilkens effects. The presence of the topological defect produced a new term on the topological nature of this effects as showed by the authors \cite{bf1} due the disclination. In this work, a new contribution is acquired due a dislocation, i. e., due the presence of the torsion.

\acknowledgments{ This work was partially supported by PRONEX/FAPESQ-PB, FINEP, CNPq and CAPES/PROCAD)}

\end{document}